\documentclass{article}

\usepackage{graphicx}
\begin{document}


\begin{center}
{\bf{MODIFICATION OF A CHARGED-BOSE-GAS MODEL FOR OBSERVED
ROOM-TEMPERATURE SUPERCONDUCTIVITY IN NARROW CHANNELS THROUGH FILMS OF
OXIDISED ATACTIC POLYPROPYLENE}}

\vskip 2em

{D.M. EAGLES}

\vskip 1em
{\em 19 Holt Road, Harold Hill, Romford, Essex RM3 8PN, England}\newline 
{\em e-mail: d.eagles@ic.ac.uk}\newline
\end{center}

\vskip 2em
[International Journal of Modern Physics B {\bf 25}, 1845-1875 (2011)]

\vskip 2em \noindent
Reasons have been found for thinking that the minimum diameter of
channels of a given length to support superconductivity at room
temperature through films of oxidised atactic polypropylene (OAPP) is
considerably larger than found in a model for Bose condensation in an
array of nanofilaments [D.M. Eagles, {\em Phil. Mag.} {\bf 85}, 1931 (2005)] 
used previously. This model was introduced to interpret experimental
results dating from 1988 on OAPP.  The channels are thought to be of
larger diameter than believed before because, for an N-S-N system where
the superconductor consists of an array of single-walled carbon
nanotubes, the resistance, for good contacts, is $R_Q/2N$, where $N$ is
the number of nanotubes and $R_Q=12.9$ k$\Omega$ [See e.g. M. Ferrier
et al.,
{\em Solid State Commun.} {\bf 131}, 615 (2004)].  We assume this would be
$2R_Q/N$ for a triplet superconductor with all spins in the same
direction and no orbital degeneracy, which may be the case
for nanofilaments in OAPP.  Hence one may infer a minimum number of
filaments for a given resistance.  In the present model, the $E(K)$
curve for the bosons is taken to be of a Bogoliubov form, but with a
less steep initial linear term in the dispersion at $T_c$ than occurs
at low $T$.  This form is different from the simple linear plus
quadratic dispersion, with a steeper initial slope, used in my 2005
paper.  A combination of theory and experimental data has been used to
find approximate constraints on parameters appearing in the theory.



\vskip 2em
\noindent
{\bf 1. Introduction}
\newline
There were claims in a 47-page deposited paper in Russian in 1988 that
superconductivity occurs at room temperature in narrow channels through
thin films of oxidized atactic polypropylene (OAPP)\cite{Gr88}, and claims 
of this in physics journals from 1989\cite{En89}.  The first 
indications of possible superconductivity at room temperature came from
experiments involving  the resistance between pressure microcontacts
on the top of OAPP films and a conducting substrate.  Three types of
contact points were found, insulating points,  points with medium resistance,
and points with low resistance.  (See Fig. 1, which is based on Fig. 17
of Ref.\cite{Gr88}, and has also been published in Ref.\cite{Ea05}).  
\begin{figure}
\rotatebox{270}
{\centerline{\includegraphics[width=9.6cm]{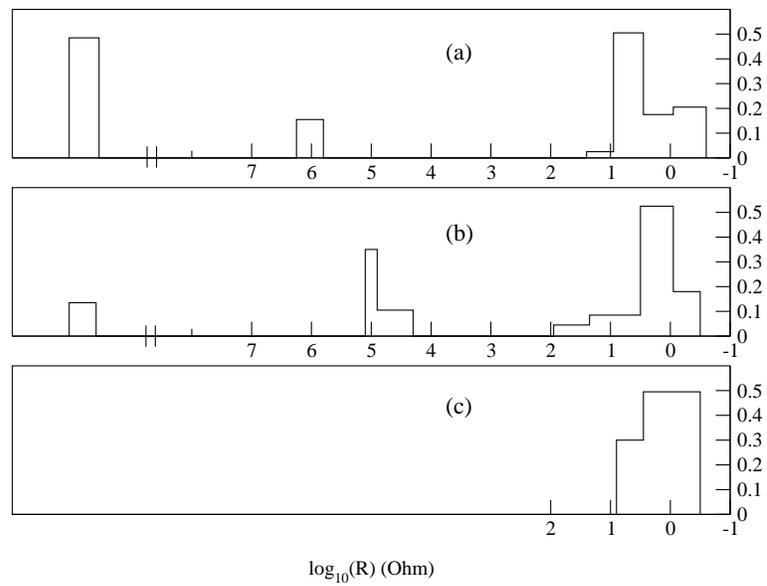}}}
\caption
{Distribution function for resistance $R$ at room temperature from
microprobes to a metallic substrate: (a) For a sample of oxidised
atactic polypropylene (OAPP) of thickness 50 $\mu$m, (b) for OAPP with
thickness 0.3 $\mu$m, and (c) for the metallic substrate without any
polymer film. [Reproduced from Fig. 1 of Ref.\cite{Ea05}, which was obtained
approximately from Fig. 17 of Ref.\cite{Gr88}]}
\end{figure}
The resistances of the group of low-resistance points lay in the same
range for two widely different film thicknesses, making it plausible
that the resistance was entirely contact resistance, with negligible
resistance in the channels between the contacts and the substrate at
these points.

More convincing evidence of room-temperature superconductivity came in
1990 and 1991 from experiments which showed (i) that the conductivity
of the low-resistance channels was several orders of magnitude larger
than that of Cu\cite{Ar90,De90}, (ii) that the high conductivity was
destroyed in pulsed measurements by non-thermal means when currents
above a critical value were applied\cite{De90}, and (iii) there was a
negligible electronic contribution to the thermal conductivity,
implying a violation of the Wiedemann-Franz law by several orders of
magnitude\cite{Gr91a}.  There are some questions in the present author's
mind about how the results of the four-probe measurements of
conductivity reported in Ref. \cite{Ar90} came about, but the estimate of
the conductivity by indirect methods in Ref. \cite{De90} seems soundly
based.  This estimate of conductivity made use of the thermal
conductivity results of Ref. \cite{Gr91a}, and depended on the lack of
decomposition of the sample after repeated pulsed measurements, which
permitted a deduction of conductivity greater than $10^{11}
\rm{ohm^{-1}cm^{-1}}$, and a transition temperature from the highly
conducting state under pulsed-current conditions of at least 700 K, the
decomposition temperature of the polymer.   Similar very high
superconducting $T_c$'s have been inferred by Zhao and coworkers 
\cite{Zh01,Zh03,Zh06,Zh08} from conductivity and other measurements on
carbon nanotubes, following a suggestion by Tsebro et al.\cite{Ts99} that
there might be superconductivity at room temperature in carbon
nanotubes.  Work claiming possible room-temperature superconductivity
in highly oriented pyrolytic graphite\cite{Ko00,Ko07} may be related.
Another claim of room-temperature superconductivity in a quasi
one-dimensional system has been made by Djurek et al.\cite{Dj01,Dj08,Dj11}.
In Djurek's work, the main material involved, $\rm{Ag_5P_2O_6}$, contains
quasi one-dimensional channels, but in recent work\cite{Dj08,Dj11} it
is found that indications of room-temperature superconductivity occur
only when pellets of $\rm{Ag_5P_2O_6}$ are embedded in an insulating
matrix.

Unusual magnetic properties have also been reported in OAPP and other 
polymers. These are:  1.  Metamagnetic transitions at magnetic fields
of the order of 1000-4000 Oersted\cite{Sm88,En89,Gr93,Gr96,Ro00} in
films of both OAPP and polydimethylsiloxane (PDMS); 2. Large
diamagnetism in some films of OAPP\cite{Gr88,En89,Ro00}; 3. Occasional
spontaneous sudden jumps of films of OAPP towards low-field regions in
inhomogeneous magnetic fields\cite{Gr93}.  The metamagnetism is
thought to be due either to a transition from an antiferromagnetic to a
ferromagnetic arrangement\cite{Gr88,Sm88,En89} or to alignment of
ferromagnetic domains of ferromagnetic (and superconducting)
filaments.  We take the second viewpoint in this paper.  The large
diamagnetism and spontaneous jumps to low-field regions in
inhomogeneous magnetic fields are thought to be associated with
superconducting channels which join at their ends to form closed loops
\cite{Gr88,En89,Gr93,Ro00,Ea02}.  In Ref. \cite{Ea02} the large diamagnetism
observed in two samples of OAPP\cite{Ro00} has been interpreted in
detail as due to induced currents in closed loops, using a model with
loops of two different radii as an approximation to a continuous
distribution.

Almost no attempts to repeat and extend the results mentioned above
have been reported in the literature.  At one point, some members of
the Ioffe Institute group who reported results supporting
superconductivity at room temperature in 1990, decided that the effects
observed were due to formation of a conducting filament of the
electrode material in high electric fields\cite{Io90,Io92}, but such
an interpretation has been shown to be untenable for the samples
studied by the group of Grigorov\cite{Gr93,Gr96}.  Also, one of the
Ioffe Institute group in more recent papers appears to have realised
that high electric fields are not needed to produce highly conducting
channels through polymer films\cite{Io00,Io08,Re08}.  Another member
of the original Ioffe group, now in Israel, has found switching to
high-conductivity states in channels through films of PDMS
\cite{Sh98}.

Baran et al. found no evidence of unusual magnetic effects in bulk atactic
polypropylene\cite{Ba95}, but in this case the oxidation was probably
not great enough, except perhaps near the surface, for conducting
channels to form.  I have been informed that a verbal report on some of the
samples of Grigorov et al. was given in the 1990's to M. Goldes (CEO of
the company with which Prof. Grigorov was associated at the time) by
P.M. Grant, who concluded that there was no superconductivity.  However,
since no details are available, the reliability of this assessment
cannot be judged.  Dr. Grant wrote a spoof essay about the disovery of
superconductivity at $T>541$ K in polymers in 2028 in Physics
Today\cite{Gra98}, but in my view put the date at about 40 years after
the real event!  Professor Grigorov appears to have stopped publication
on the subject since 2000, concentrating instead on improving samples
etc., and patent work\cite{Gr98,Gr03,Gr06}, and so, unless
another group takes up studies in the field, it may be a long time
before his work is properly assessed.

Only one group, originally at the Institute of Synthetic Polymeric
Materials in Moscow, has consistently claimed room-temperature
superconductivity in channels through films of OAPP, PDMS and other
polymers, including polyoctylmethacrilate\cite{Kr99}. However, an
early paper from the Ioffe Institute, and also more recent work by some
scientists from the Ioffe Institute, especially Ionov, who was a member
of the original group publishing in 1990, collaborating  with others
from Russia and Germany, have found evidence that superconductivity
occurs in narrow channels through many polymer films at temperatures
below the $T_c$ of superconducting electrodes, and that this is not due
to the superconducting proximity effect.  Such evidence includes
conductivity greater than $10^{14}\;\Omega^{-1}$cm$^{-1}$ in channels
through films between superconducting Sn electrodes\cite{Ar90},
Josephson-like I-V curves for
superconductor-polymer-superconductor sandwiches in OAPP\cite{Ar90},
PDMS\cite{Io00},  poly(phthalidylidene biphenyline) \cite{Io02a},
polyamidine\cite{Io08}, and two other polymers\cite{Io02b},
oscillations of resistance as a function of magnetic field in films of
polyimide, poly(phthalidylidene biphenylene) and PDMS below the $T_c$
of Sn\cite{Io99,Io00,Io02a}, and a critical current for a destruction
of Josephson-like $I-V$ curves which depends on temperature in a way
inconsistent with superconductivity due to the proximity effect in PDMS
\cite{Io00} and poly(phthalidylidene biphenylene)\cite
{Io02a,Io09a}.  There is also a large number of papers by the group of A.N.
Lachinov in Ufa on highly conducting channels through polymer films
(see e.g.\cite{La93,Sc95} or a review \cite{La06}), but, except in
recent work with Ionov and others\cite{Za98}, without suggestions
of superconductivity.  The work from 1998 onwards sometimes made use of
spin-coated films, which method of production allows better control of
film thickness\cite{Za98}.  Recent work of Ionov and Rentzsch\cite{Io10a,Io10b}
considers the possibility of resonant tunnelling to
explain their results.  A difference between results of Ionov et al and
those of Grigorov's group is that Ionov and coworkers report a
dependence of conductance of
channels on the work function of the substrate\cite{Io09b}, whereas no
such dependence is reported by Grigorov et al..   Possibly the
difference is related to how good the contacts with the substrate
are.   A detailed discussion of different forms of $I-V$ curves for
different types of contacts to OAPP and PDMS
films is given by Kraev et al.\cite{Kr93}.

A theory for why conducting channels exist in films of elastomers
(materials with very low elastic constants) has been developed by
Grigorov  and coworkers\cite{Gr90a,Gr90b,Gr91b}.  The  theory involves
the formation of one-dimensional channels with diameters of the order of
nanometers due to the formation of "superpolarons", i.e.
one-dimensional strings of a certain type of polaron involving electrons
interacting with rotatable dipolar groups.   An unusual theory for
superconductivity in such channels involving motion of oppositely
directed charge density waves was developed by
Grigorov\cite{Gr98b,Gr99}.  An earlier theory by me for the
superconductivity involving large enhancements of pairing interactions
at certain high drift velocities\cite{Ea94,Ea98a} has had to be abandoned,
because recent work on bipolarons in one dimension\cite{Ea07} gives no
support for the greatly enhanced interactions at such velocities predicted by 
perturbation-type methods.  However, work\cite{Ea05} involving
Bose condensation of bipolarons (without any special enhancement of
interactions at high drift velocities),  with an initial linear
dispersion at low wave vectors may be on the right track, and the
purpose of this paper is to improve on the model introduced in 2005,
making more use of both experiment and theory to help to determine
parameters in the theory.

The model and constraints we use differ in four ways from that of
2005.  First the minimum number of nanofilaments required to support
superconductivity at room temperature is taken to be considerably
higher than in the 2005 paper.  This is because we have realised that,
in the case of superconducting channels consisting of single-walled
nanotubes between normal-metal electrodes, there is a
minimum resistance equal to $R_Q/2N$ for singlet superconductivity, where
$N$ is the number of nanotubes\cite{Fe04}, and $R_Q=12.9$k$\Omega$ is
the resistance quantum.  For nanotubes there is an orbital degeneracy
as well as the spin degeneracy, which gives rise to the factor of two
in the denominator of the above expression.   For nanofilaments  with
no orbital degeneracy and triplet superconductivity with all spins in
the same direction, as is thought to be the case for OAPP, we assume
that $R_Q/2N$ becomes $2R_Q/N$, where $N$ is the number of
nanofilaments. 

The second difference from 2005 arises since we doubt if the form
of dispersion for bosons assumed in 2005 based on a Cooper-pair
model\cite{Ad00} is appropriate in the Bose-gas regime, where the
Fermi surface no longer exists even above $T_c$.   We assume instead
that the dispersion has a Bogoliubov-type form, but with the strength
of the initial linear slope at $T_c$ (related to an effective bosonic
chemical potential) smaller than at low $T$.  This hypothesis is made
as an attempt to account for the fractional increase in $T_c$ of an
interacting Bose gas due to interactions, proportional to $n_b^{1/3}$
for low boson concentrations $n_b$ \cite{Ar01,Ka01}, in a model of
non-interacting bosons with a different type of dispersion from the
bare one, in a similar spirit to Fermi-liquid theory.  At higher boson
concentrations the fractional change of $T_c$ of an interacting Bose gas
due to interactions decreases with $n_b$\cite{Pi08}.  This can be taken
into account in a non-interacting boson model by a mass which increases
with $n_b$, but with a dependence on $n_b$ with a higher power than 1/3.
Although there is no {\em a priori} reason to choose a Bogoliubov-type
of dispersion at $T_c$, a possible justification is that above $T_c$
there are fluctuating regions of superfluid which support this type of
dispersion locally, in analogy to the Aslamozov-Larkin theory\cite{Az68}
for BCS superconductors.  The steepness of the initial linear term,
proportional to the square root of the effective bosonic chemical
potential, is chosen to give approximate agreement of $T_c$ for the
non-interacting bosons with a Bogoliulobov-type of dispersion with
numerical calculations mentioned above for an interacting Bose gas at low
boson concentrations.  Although the Bogoliubov dispersion is modified
in a charged Bose gas (see e.g.\cite{Al93}), we do not expect great
changes from the simple Bogoliubov form at temperatures  close to the
condensation temperature because
screening will cause the interactions to become short range at boson
concentrations of interest.

The third difference from 2005 is that we make more use of theoretical
work (see especially\cite{Gr91b}) to put constraints on the many
parameters in the theory.  We determine parameters by minimising a
weighted sum of moduli of differences of two sides of equations given
by constraints, with the weighting varied according to how accurate we
think the constraints are - several of these are thought to be at best
only of order-of-magnitude accuracy.  We estimate the linear carrier
concentration for the 50 $\mu$m film from the theory of 
Grigorov\cite{Gr91b} which minimises the energy of a ferromagnetic
nanofilament,  but, in the 0.3 $\mu$m film we use the linear carrier
concentration as an adjustable parameter because charge injection from
the electrodes may influence the carrier concentration for
films of thicknesses less than about 3 $\mu$m\cite{Io05} (see also
Ref. \cite{Lo80},  although the thickness to which this may occur does not
appear to be discussed in this reference), and this injection may
either increase or decrease the total electron concentration.

There are twelve parameters which need to be determined in our theory.
These are: 1) and 2)  The numbers $n_T$ of nanofilaments in each
transverse direction of assumed square arrays of nanofilaments for the
highest-resistance superconducting channels for 50 and 0.3 $\mu$m films;
3) The lattice constant $a_T$ of these arrays; 4)  and 5) The product
$c_Ba_T$ of the linear boson concentration $c_B$ in a nanofilament with
this lattice constant in 50 $\mu$m films, and the change $\Delta c_Ba_T$
of this product in the 0.3 $\mu$m films because of possible effects of
carrier injection from electrodes; 6) The ratio of $a_T/\rho$ of $a_T$
to the radius $\rho$ of the cylinder around a nanofilament where all
dipolar groups are pointing away from the axis of the nanofilament; 7)
and 8) The binding energy $E_{pb}$ and the longitudunal mass $M_L$ of the
pairs which form in the polaron string due to mediation of
high-frequency excitations; 9) The ratio $p=M_L/M_T$ of the
longitudinal and transverse masses $M_L$ and $M_T$ of bosons in a
nanofilament array; 10) The energy of the excitation which mediates the
pairing attraction between the polarons within the superpolaron array,
which excitations we assume for most of the paper to be plasmons; 11)
The effective bosonic chemical potential $\mu_B$ which appears in the
assumed Bogoliubov form of the boson dispersion at $T_c$; and 12) The
reduction $\Delta T_1$ of the superconducting transition temperature for
the highest-resistance superconducting channel for the 50 $\mu$m films
due to phase slips.  We do not have a method of determining the last
parameter without a detailed theory of phase-slip resistance for wires
with BEC-type superconductivity, but, assuming $T_c$ remains above room
temperature for all cross sections larger than the minimum required to
support superconductivity at that temperature, we can determine a lower
limit on $\Delta T_1\approx 7$ K within the framework of our model.  We
do most of our calculations with $\Delta T_1$ = 15 K, not far above the
lower limit of about 7 K.

After writing down the basic equations of our model in section 2, in
section 3 we find  values for the first two parameters from Fig. 1
assuming good contacts.  Then, from a combination of theory and use of
experimental data, sections 2 and 4 to 11 enable us to find nine
equations of varying degrees of approximation to be satisfied by
parameters 3), 4), and 6) to 11), assuming  a value of  15 K for
$\Delta T_1$.  In section 12 we find the change of $c_Ba_T$ for the
thinner films to make $T_c$ equal to room temperature for the $n_T$ for
the highest-resistance superconducting channel for these films.

Some of the equations are thought to be fairly accurate while others
are at best order-of-magnitude constraints.   For our numerical work to
determine parameters in section 13 we weight the various constraints
according to our guesses at their accuracy.
Some discussion is given in section 14.   

\vskip 2em
\noindent
{\bf 2.  Model of Non-interacting  Bosons
in an Array of Nanofilaments with a Bogoliubov Type of Dispersion at $T_c$}
\newline
As in Ref. \cite{Ea05}, we use a model of conducting channels composed of 
non-interacting bosons on a square lattice forming a square
array of quasi
one-dimensional filaments  of length $L$ and transverse dimension $D$
measured from a point half a filament separation $a_T$ beyond the centers
of the edge filament rows.  The filaments are assumed to be
sufficiently narrow that the bosons in a filament are
in the lowest possible state with respect to motion perpendicular to
the filaments at all temperatures of interest.    We suppose that boson
wave functions vanish at the edges of the array, and neglect any
periodicity in the potential along the length of the array.  Then we
suppose that the boson states in the array can be characterised by the
magnitudes of wave numbers $K_1$, $K_2$ and $K_3$ in the two transverse
and longitudinal directions, where $K_1$ and $K_2$ take on discrete
values ranging from $\pi/D$ to $n_T(\pi/D)$, and $K_3$ takes on
values of $k\pi/L$, where $k$ runs from 1 to $\infty$.     
For computational purposes we approximate the sum to infinity by a finite
sum with a suitably high upper limit. 

The difference in our model from that used in Ref. \cite{Ea05} is in the form
of the $E_{\bf K}$ chosen for the bosons.  Here we assume that the
boson energy $E_B(K_1,K_2,K_3)$, where $K_1=i\pi/D$, $K_2=j\pi/D$,  
$K_3=k\pi/L$, has the Bogoliubov form 
\begin{eqnarray}
E_B(K_1,K_2,K_3)=&\nonumber\\
\mbox{\large [}\{A_L(\pi/L)^2&[(k^2-1)+p(L/D)^2(i^2+j^2-2)]+\mu_B\}^2-\mu_B^2
{\mbox\large]}^{\frac{1}{2}}.
\end{eqnarray}
Here 
\begin{equation}
A_L=\hbar^2/2M_L,
\end{equation}
and
\begin{equation}
p=M_L/M_T,
\end{equation}
where $M_L$ and $M_T$ are the boson masses in the longitudinal and
transverse directions; $\mu_B$ is an effective chemical potential of
the bosons at the temperature being considered.  We say an effective
bosonic chemical potential because the actual bosonic chemical
potential at the Bose-condensation temperature must be zero, but, as
mentioned in the Introduction, we suppose that interaction between
bosons may be taken into account by a Bogoliubov form of dispersion for
non-interacting bosons which give about the same type of increase in $T_c$ as
accurate theories of the effect of interactions find.  The  same type
of model with a simple quadratic dispersion was discussed in
Ref. \cite{Ea98b}, but with the difference from the case assumed here that a
discrete lattice was considered in the longitudinal as well as
transverse directions.

Condensation into the lowest-energy state cannot take place if the sum
of the boson occupation factors over all states except the lowest for a
chemical potential situated at the energy of the lowest state is
greater than the total number of bosons, $N_B$.  Below the temperature
at which this sum is equal to $N_B$, the occupation of the ground state
and of states in its neighbourhood will start to become macroscopic,
i.e. of the order of $N_B$, and so the equality of the sum with $N_B$
determines the condensation temperature $T_B$. 
For an average linear concentration $c_B$ of bosons per filament,
this criterion can be written as
\begin{equation}
\sum_{K_1,K_2,K_3}\frac{1}
{{\rm exp}[E_B(K_1,K_2,K_3)/k_BT_B]-1}
= c_BL n_T^2,
\end{equation}
where the sums over $K_1$ and $K_2$ are sums from $i$, $j$ = 1 to $n_T$ of
$K_1=i\pi/D$ and $K_2=j\pi/D$, and the sum over $K_3$ is a sum from $k=1$
to infinity of $K_3=k\pi/L$, but the ground level $i=j=k=1$ is excluded.

As discussed in Ref. \cite{Ea05}, it is difficult to define unambigously a
condensation temperature for small arrays because of the gradual way in which
occupation numbers of the ground state change with temperature for such arrays
\cite{Al01,Al02}.  However, we shall be concerned here mainly with arrays
containing more than about 10,000 bosons, and so this ambiguity should
not be a problem.

\vskip 2em
\noindent
{\bf 3.  Constraints from Histograms of Resistances
with Microcontacts}
\newline
Because of the lack of dependence of the average resistance on film
thickness of the lowest-resistance parts of the histograms in Fig. 1,
and for other reasons discussed in the Introduction, we assume that the
channels associated with these parts of the histograms are
superconducting at room temperature.  Further, based on previous
studies\cite{Ea98b,Ea05} which found that, at least for fairly low
cross sections of channels, the Bose condensation temperature of
nanofilament arrays increases with the array cross section, we suppose
that the superconducting $T_c$ tends to increase as the resistance 
measured by the two-point probe method decreases.  Hence we assume that
the highest-resistance channels of the  superconducting group (HRSC's)
have superconducting $T_c$'s of just about room temperature, i.e. we
assume that 
\begin{equation} 
T_{c,HRSC}(L) \approx 295 K, 
\end{equation}
where $L$=50 $\mu$m or 0.3 $\mu$m for the highest-resistance channels of
the low-resistance groups in Figs 1(a) and (b) respectively.

For an array of single-walled superconducting carbon nanotubes
with good contacts from normal metals,
there is a minimum resistance\cite{Fe04} 
\begin{equation}
R_m = R_Q/2N, 
\end{equation}
where 
\begin{equation}
R_Q=12.9 \rm k\Omega
\end{equation}
is a resistance quantum.  Single-walled carbon nanotubes have two
orbitally-degenerate and spin-degenerate subbands\cite{Fe04,Sch99}.  
We assume that this
minimum resistance becomes $2R_Q/N$ for triplet superconductivity in
nanofilaments with all the spins of a nanofilament in the same
direction, as is thought to be the case for OAPP
\cite{Gr90a,Gr90b,Gr91b}.  We have no reason to assume orbitally-degenerate
subbands, and, if all spins are aligned, this also halves the
number of current paths.  Hence, assuming good contacts, we suppose that
$n_T$ for the highest-resistance superconducting channels for the two
film thicknesses being considered satisfies
\begin{equation}
n_{T,HRSC}(L)^2 \approx 2R_Q/R,
\end{equation}
where $R$ is the resistance for the relevant channel.
Using data from Fig. 1 we deduce 
\begin{equation}
n_{T,HRSC}(0.3 \mu{\rm m}) \approx 17,\;n_{T,HRSC}(50 \mu{\rm m}) \approx 32.
\end{equation}

For samples with small cross sections, phase slips cause 
resistance\cite{La67,McC70}.
Defining, arbitrarily, $T_c$ of the system as that temperature at which
phase slips cause a resistance of half the resistance in the normal state,
then phase slips reduce an initial superconducting transition temperature
$T_{c0}$ to $T_c$,
with 
\begin{equation}
T_c=T_{c0}-\Delta T
\end{equation}
where, for BCS-type superconductors\cite{La67},
\begin{equation}
\Delta T \propto \rm{(cross}-\rm{section)}^{-2/3} \propto n_T^{-4/3}.
\end{equation}
Although there are papers on phase slips for Bose condensates
\cite{Po05a,Po05b}, they do not appear to give sufficient information
to determine the dependence of $\Delta T$ on $n_T$ or on coherence
lengths for such systems.   Thus we use Eq. (11) to find the dependence
of $\Delta T$ on $n_T$ for a given boson concentration in our numerical
work.

\vskip 2em
\noindent
{\bf 4.  Relation between Linear Boson
Concentration and Other Parameters based on Grigorov's 1991 theory}
\newline
In an elastomer (a material with very low elastic moduli) containing rotatable
groups of atoms with electric dipole moments, an excess electron causes
all dipoles to point
away from the electron within a radius $R_0$ given by\cite{Gr90a}
\begin{equation}
R_0 = (eD_m/\pi v G \epsilon_h)^{\frac{1}{2}},
\end{equation}
forming an unusual type of polaron.  Here $D_m$ is the dipole moment of a group,
$v$ is its volume, $G$ is a shear modulus, and $\epsilon_h$ is a 
high-frequency dielectric constant.  For
many electrons in such a material it is energetically favourable for
the polarons to form strings called "superpolarons"
\cite{Gr90a,Gr90b,Gr91b}.  For superpolarons with a linear concentration
of polarons $c_F$, the dipoles point outward in a cylinder of radius
$\rho= 2 R_0^2 c_F$\cite{Gr91b}.  In our theory we suppose that there
is pairing of polarons within a string caused by attraction mediated by
plasmons or high-energy phonons, to give a linear concentration of
bosons $c_B=0.5c_F$, and so, putting results in terms of boson
concentrations we have
\begin{equation}
\rho=4 R_0^2 c_B. 
\end{equation}
When exchange forces are included, Grigorov\cite{Gr91b} finds that, for
low carrier concentrations, the minimum energy of a polaron string occurs when
the polarons are ferromagnetically aligned with a linear concentration such
that
\begin{equation}
k_FR_0 \approx 1.6,
\end{equation}
where $k_F$, the Fermi wave number for spin-aligned electrons, is
\begin{equation}
k_F = \pi c_F = 2\pi c_B.
\end{equation}
Ferromagnetism at low concentrations in a quasi one-dimensional
electron gas has also been discussed by Shelykh et al.\cite{Sh03}.

We consider a square array of polaron strings (to be referred to as
nanofilaments from now on) with lattice constant $a_T$.  From Eqs. (13-15), 
we find that 
\begin{equation}
c_B a_T = 4 \times (1.6/2\pi)^2 (a_T/\rho) = 0.259 (a_T/\rho).
\end{equation}

\vskip 2em
\noindent
{\bf 5.  Relation between
Separation of Nanofilaments and Radius of Fully Aligned Dipole
Region from Experimentally Based Estimates of Average Carrier
Concentration in Conducting Channels}
\newline
The average concentration of bosons within channels is not known
accurately.  However, we may obtain an order-of magnitude estimate of
this as follows.  From magnetisation results, the maximum concentration
of aligned spins in any sample reported on up till now appears to be that
for sample 1 of Ref. \cite{Ro00}, where the saturation magnetisation after
any diamagnetic contribution has been removed by field cycling is 0.058
emu cm$^{-3}$ (see dotted curve in Fig. 1 of Ref. \cite{Ro00}).  Hence,
assuming this moment is due to aligned electron spins, the average
concentraion $n_F$ of fermions in this sample is given by
\begin{equation}
n_F\approx 6.3\times 10^{18}\rm{cm}^{-3}.  
\end{equation}
The highest concentration of conducting channels reported in any sample
is given in Ref. \cite{Sm89}, where the average separation of the channels
was determined to be 7.5 $\mu$m.  Hence with a typical channel diameter
of about\cite{De90} 1 $\mu$m, the ratio $V_s/V_c$ of the volume of the
sample to the total volume occupied by conducting channels is  given by
\begin{equation}
V_s/V_c \sim 7.5^2\times (4/\pi) = 72.
\end{equation}
If we suppose that the fraction of material occupied by channels in sample 1 
of Ref. \cite{Ro00} is about the same as in the sample of Ref. 
\cite{Sm89}, then,
from Eqs. (17) and (18), we deduce that the concentration of bosons
$n_B$ in the channels (half the fermion concentration) satisfies
\begin{equation}
n_B \sim 2.2 \times 10^{20} \rm{cm}^{-3}.
\end{equation}
Using 
\begin{equation}
n_B = (c_Ba_T)a_T^{-3}
\end{equation}
and Eqs. (16) and (19), we deduce that
\begin{equation}
a_T \approx 1.04(a_T/\rho)^{1/3} {\rm nm}.
\end{equation}
Although we used $\sim$ in Eq. (19), we have
changed to $\approx$ in Eq. (21) because of the one-third power.

\vskip 2em
\noindent
{\bf 6.  Estimate of Transverse
Bipolaron Mass based on Grigorov's 1991 Theory}
\newline
In this section we make use of the theory of Ref. \cite{Gr91b} to
estimate the transverse mass for motion of particles across our array
before inclusion of pairing due to mediation of plasmons or
high-frequency phonons, and use these calculations to estimate the
ratio $M_L/M_T$ of longitudinal and transverse bipolaron masses in
terms of other parameters.  Assuming a tight-binding model for
transverse motion across the array, the transverse single-particle
masses can be found in terms of energy overlap integrals $t$.

We suppose that the diameter $D$ of the neutral cylinder enclosing a
polaron string in the theory of Ref. \cite{Gr91b} can be identified with
$a_T$ for our array of strings.  The potential energy appearing in an
expression for $t$ has to be measured from a zero equal to the lowest
energy state of an isolated string.  This
energy, which we call $E_0$, is given by
\begin{equation}
E_0 = 2\pi\int_{0}^{\infty} [\chi^2(r) V(r) - (\hbar^2/2m_e)\chi d^2 
\chi/dr^2] dr, 
\end{equation}
where the (real) radial wave function $\chi(r)$ is given by
\begin{equation}
\chi(r) = (1/2\pi)^{1/2}(2/3) 2^{1/2} \beta(1+\beta r) \rm{exp}(-\beta r),
\end{equation}
with
\begin{equation}
\beta = 5/R_0 =5 \times 2\pi c_B/1.6 =19.6 c_B
\end{equation}
in our notation.   We have changed slightly from Grigorov's notation
in that we have included a factor $(1/2\pi)^{1/2}$ in $\chi$.
The potential energy of an electron $V(r)$ in our notation satisfies
\begin{eqnarray}
V(r)=
 -[e^2(1-k)/(\epsilon_h a_T)]\nonumber\\ 
\{2c_Ba_T[2{\rm ln}(a_T/2\rho)+(2\rho/a_T)^2-1]+a_T(\rho-r)/R_0^2\}
\;\; (r<\rho)\nonumber\\
= -[2e^2(1-k)c_Ba_T/(\epsilon_h a_T)] [2{\rm ln}(a_T/2r)+(2r/a_T)^2-1] \;\; 
(\rho<r<a_T/2)\nonumber\\
=0\;\; (a_T/2<r).
\end{eqnarray}
Here 
\begin{equation}
k=\epsilon_h/\epsilon_s, 
\end{equation}
where $\epsilon_s$ is the static dielectric constant, and $\epsilon_h$,
is the high-frequency dielectric constant.   For materials in which
polaron strings occur, $k$ is small compared with unity.

For wave functions as given by Eq. (23), we find that the 
energy overlap integral $t$ is given
by 
\begin{equation}
t = \int_{0}^{\infty}\int_{0}^{2\pi} 2\pi r [V(r)-E_0)]\chi(r) \chi(u) 
dr d\theta,
\end{equation}
with
\begin{equation}
u = [a_T^2+r^2 -2a_T r \rm{cos}\theta]^{1/2}.
\end{equation}

The bare transverse mass $m_{FTb}$ of fermions for our square lattice in a
tight-binding model is related to the energy overlap integral $t$ by  
\begin{equation}
m_{FTb} = \hbar^2/2ta_T^2.
\end{equation}
Here we use bare in the sense of not including any effects of interactions
with high-frequency modes which give rise to the pairing in our model.

For a Holstein model with anisotropy\cite{Al08}, the ratio of the
transverse to longitudinal polaron masses is not much different from the
same ratio for the bare masses, at least for interactions with
high-frequency phonons.  Thus, combining this result with the
assumption that the  ratio of the pair transverse mass $M_T$ to the
bare transverse fermion mass $m_{FT}$ is approximately equal to the
ratio of the pair longitudinal mass $M_L$ to the bare fermion longitudinal
mass, which we take as equal to the free-electron mass $m_e$, we find
\begin{equation}
M_T \approx 2m_{FTb} (M_L/2m_e).
\end{equation}

\vskip 2em
\noindent
{\bf 7.  Estimate of Energy of Mediating Bosons 
assuming these are Plasmons} 
\newline
We assume for most of this paper that the bosons mediating the
attraction between fermions in the nanofilaments are plasmons, although
we make some remarks in the discussion section about the alternative
that they are high-frequency phonons.

As in Ref. \cite{Ea94}, we estimate the plasmon energy as that for a single
nanofilament at wave numbers $q$ such that $qd\stackrel{>}{\sim} 1$,
where $d$ is the diameter of the nanofilament.   In this case the
plasmon energy is
\begin{equation}
\hbar \omega = \hbar(16\pi e^2 c_Bd^{-2}/M_L\epsilon_h)^{1/2},
\end{equation}
where $c_B$ is the linear boson concentration  and $\epsilon_h$ is the
high-frequency dielectric constant.  The difference from the usual factor
of $4\pi$ in the bracket comes about since the 
pair charge is $2e$.
For the diameter $d$ in Eq. (31), we first estimate the mean value of
the radius from the axis from Grigorov's variational method\cite{Gr91b} as
\begin{equation}
r_m=\int_0^{\infty}r^2(1+\beta r)^2e^{-2\beta r}/
\int_0^{\infty}r(1+\beta r)^2e^{-2\beta r}=(14/9)\beta^{-1},
\end{equation}
where, from Eq. (24), $\beta=19.6(c_Ba_T)/a_T$.   For a uniform
distribution of charge up to a cylinder radius of $r_c$, the mean value
$r_m$ of $r$ is given by
\begin{equation}
r_m=(2/3)r_c.
\end{equation}
Thus to convert the mean value of $r$ given above to a total cross
section, we multiply the cross section by (9/4), or the linear
dimensions by (3/2).  Hence the diameter of an equivalent square
filament is taken as
\begin{equation}
d=(3/2)(14\pi^{1/2}/9\beta)=0.211a_T/(c_Ba_T).
\end{equation} 
Hence, from Eq. (31), 
\begin{equation}
\hbar\omega\approx\hbar(16\pi e^2c_Ba_T/\epsilon_h M_La_T^3)^{1/2} 
(4.74c_Ba_T).
\end{equation}
There will be additional terms proportional to the square of the wave
number and inversely proportional to the mass at high $q$, but, for
parameters of interest, these are not expected to change the relevant
average plasmon energy by large amounts, and will be partially
cancelled by the reduction in plasmon energy at small $q$.  

\vskip 2em
\noindent
{\bf 8.  Approximate Relation between the Bipolaronic Binding Energy
and the Bipolaron Mass}
\newline
We obtain another approximate relation between parameters by making use
of results for bipolaron binding energies and masses for weak to
intermediate coupling in a one-dimensional system\cite{Ea07}.  Although
these results were obtained with symmetric spatial bipolaron wave
functions, implying singlet bipolarons, results for singlet and triplet
bipolarons in a Holstein-Hubbard model do not look drastically
different from each other for weak to intermediate coupling (see Figs.
1 and 2 of Ref. \cite{Ha07} or Fig. 15 of Ref. \cite{Ha09}). 
From section 4 of
Ref. \cite{Ea07}, for $V^2$ = 0.125, 0.25 and 0.375 and $a=4$ in the
notation of that paper, $a=4$ corresponding to high frequencies of the
bosons mediating the electron-electron attraction, we can see that the
ratio $(E_b/\hbar \omega)/[(M_{bp}/2m_b)-1]$ lies between 1.63 and
1.32,  and, with extrapolation to higher $V^2a$ up to 0.5, this value
could go down to about 1.25.  For values of $V^2$ between 0.125 and 0.375, the
values of the bipolaron binding energies $E_{pb}$ lie between
0.08$\hbar\omega$ and 0.38$\hbar\omega$, while the bipolaron masses
$M_L$ (for a bare mass $m_b$) lie between 2.13 and 2.5 $m_b$  for the
same parameter range.  Based on these results we shall suppose that
\begin{equation} 
M_L/m_b = 2.0+(1.4\pm 0.2) (E_{pb}/\hbar\omega), 
\end{equation} 
and in our case we suppose $m_b=m_e$,
the free-electron mass.

$V$ and $a$ in our 2007 model, which uses local electron-boson
interactions but a quadratic dispersion for bare electrons, correspond
approximately to what are usually called $g$ and ($\hbar
\omega/t)^{1/2}$ in the Holstein model.  Thus $0.5V^2a^2$ corresponds
to $g^2\hbar \omega/2t$, usually called $\lambda$ in the 1-D Holstein
model.  Hence $V^2a = 0.5$ and $a=4$ imply $\lambda \approx 2$.  Fig. 1
of a paper by Hohenadler and von der Linden\cite{Ho05} shows that the
kinetic energy (related to reciprocal masses) of Holstein bipolarons is
reduced by a factor of more than five at $\lambda$=2.  However, we
expect reductions of masses in our model compared with those on the
Holstein model to become larger as the coupling increases, and so the
comparatively small masses for the values of $V^2$ and $a$ with which
we have been concerned may be realistic for our model.

\vskip 2em
\noindent
{\bf 9.  Probable Lower and Upper
Limits for the Bipolaron Binding Energy}
\newline
Although there are criteria for being on the BEC side of the BEC to BCS
transition in terms of the product of the scattering length and the
Fermi wave vector\cite{Pi05,Ch05,Je06} or in terms of the ratio of
$E_{pb}$ to $E_F$ (deducible from Fig. 1 of Ref. \cite{Ea69}), we are not
sure how these are modified for systems which are both polarised and in
which there is also a cut off in wave vector in two directions, with the
Fermi wave vector having a value which is approximately equal to its 1-D
value.  (How good this approximation is will be discussed after
numerical work determined approximate values of parameters).

The simplest criterion for being on the BEC side of the transition is
that the diameter of the pairs is smaller than their mean separation.
In the longitudinal direction, the diameter of the pairs is
approximately $(8\hbar^2/M_L E_{pb})^{1/2}$, assuming $M_L \approx$ (twice
the longitudinal fermion mass), and the pair separation is approximately
$(c_Ba_T)^{-1}a_T$.  Hence a lower limit for the pair binding energy is
\begin{equation}
E_{pb}=16(\hbar^2/2M_La_T^2)(c_Ba_T)^2.
\end{equation}

Another lower limit on the pair binding energy requires that $E_{pb}$ is
appreciably larger than 3.5 $k_B\times$ 295 K $\approx 0.08$ eV (BCS
value of energy gap for $T_c=295 K$).  We use the first of these limits,
but after numerical work we also check that the value found is also
above the second lower limit.  

For an upper limit, based on our discussion in the previous section,
and the assumption that we do not wish to venture far out of the
intermediate-coupling r\'egime, we suppose that the pair binding energy
$E_{pb}$ satisfies $E_{pb} <0.6 \hbar \omega$.  Hence, using Eq. (37), we
put
\begin{equation}
E_{pb} \sim 0.5[16\hbar^2/(2M_L a_T^2)(c_Ba_T)^2+0.6\hbar \omega]. 
\end{equation}

\vskip 2em
\noindent
{\bf 10. Order of Magnitude
Constraint from Critical Current Density}
\newline
Estimates of the critical current density made from pulsed-current
measurements vary from\cite{De90b} $10^8$ Acm$^{-2}$ to\cite{De90} 
$5\times 10^9$Acm$^{-2}$, depending on whether  the size of the pressure
contacts or the estimated cross section of a single channel is used to
convert a critical current of ($63\pm 17)$ A, for gradually increasing
magnitude of the currents in the pulses, to a current density.  We take
the geometric mean of these two types of estimates, i.e.
\begin{equation}
j_c \sim 7 \times 10^8 {\rm A cm}^{-2}.
\end{equation}

In numerical studies of a fermion superfluid, Heiselberg\cite{He06}
shows that, for  the Bose-gas regime, the sound velocity at low
temperatures (equal to the critical drift velocity $v_d$  for a Bose
gas) lies between  about 0.08 and 0.25 times the Fermi velocity,
$v_F$.  Thus we assume that
\begin{equation}
v_d=(0.165\pm 0.085) v_F. 
\end{equation} 
Since the critical temperature in the pulsed currents used is estimated
in Ref. \cite{De90} to be greater than 700 K, we assume that room
temperature will count as a fairly low temperature for the purposes of
estimating the critical current.  Hence, using Eqs. (39) and (40), we
find for our model that
\begin{equation} 
(0.33\pm 0.17) e(c_Ba_T)a_T^{-3}v_F\sim 7\times
10^8{\rm Acm}^{-2} = 2.1 \times 10^{18} \rm{esu}.  
\end{equation} 
We approximate the Fermi velocity by that for a one-dimensional
ferromagnetic system as in the previous section.  Then we find
\begin{equation} 
v_F\approx\hbar (2\pi)c_B/m_{FL} \approx
\hbar(4\pi)(c_Ba_T)/a_TM_L.  
\end{equation} 
Eqs. (41) and (42) give
another order-of-magnitude relation between parameters of our model.

\vskip 2em
\noindent
{\bf 11.  Estimate of Effective Pair Chemical Potential} 
\newline
We estimate an effective pair chemical potential to put into the
Bogoliubov-type of dispersion for non-interacting bosons at $T_c$ as
follows.  First we look at accurate results for the condensation
temperature for interacting bosons\cite{Ar01,Ka01} at low carrier
concentrations, where it is found that the change $\Delta T_c$ from the
non-interacting boson value $T_{c0}$ is given by
\begin{equation}
\Delta T_c/T_{c0} \approx 1.3 a_B n_B^{1/3},
\end{equation}
where $a_B$ is the boson scattering length.  
Equation (43) is only valid for low boson concentrations, but decreases of
$T_c$ due to interactions for higher concentrations\cite{Pi08} can
be taken into account in a non-interacting boson model by assuming an
appropriate increase in mass with $n_B$ with a dependence on density
with a higher power of $n_B$ than 1/3.

The boson scattering length in Eq. (43) satisfies
\begin{equation}
a_B \approx 0.6 a_F,
\end{equation}
where $a_F$ is the fermion scattering length\cite{Pe04},
which in turn is related to the pair binding
energy $E_{pb}$  and fermion mass $m_F$ by\cite{Pi04}
\begin{equation}
\hbar^2/(m_F a_F^2)=E_{pb}.
\end{equation}

Assuming the fermion mass is approximately equal to half the boson
mass $M_B$, we deduce  from Eqs. (44) and (45) that
\begin{equation}
a_B \approx 0.6(4\hbar^2/2M_B E_{pb})^{1/2}.
\end{equation}
For an anisotropic system we suppose that $a_B$ to use in (43)
is as in (46), but with
\begin{equation}
1/M_B^{1/2} \rightarrow (1/3)(1+2p^{1/2})/M_L^{1/2}
\end{equation}
in our notation.  Also, in our system,
\begin{equation}
n_B^{1/3}=(c_Ba_T)^{1/3} a_T^{-1}.
\end{equation}
Hence, from Eqs. (43) to (48) we find
\begin{equation}
\Delta T_c/T_{c0} \approx 0.52(\hbar^2/2M_L E_ba_T^2)^{1/2}(1+2p^{1/2})
(c_Ba_T)^{1/3}
\end{equation}

For non-interacting bosons with a Bogoliubov type of dispersion at $T_c$,
we find numerically the $\mu_B$ which gives the same fractional change of 
$T_c$ from that for a pure quadratic dispersion as for the interacting system
as determined above.  Hence we find $\mu_B$ as a function
of our other parameters.  

\vskip 2em
\noindent
{\bf 12.  Change in Carrier Concentration in 0.3 $\mu$m Films required
to make the $T_c$ of the Highest-Resistance Superconducting Channel
for these Films equal to Room Temperature}
\newline
Now, because of possible injection of carriers from the electrodes (see
the Introduction), we assume that
\begin{equation}
c_Ba_T \rightarrow c_Ba_T+\Delta (c_Ba_T) = c_{B2}a_T \;\rm(say)
\end{equation}
for 0.3 $\mu$m films.  Then $p$ in our equations in section 2 has to be
replaced by $p_2=M_L/M_{T2}$, where the transverse mass $M_{T2}$ for the 
thinner films is calculated as in section 6, but with $c_Ba_T$ replaced by
$c_{B2}a_T$.

For a fixed boson concentration, the decrease $\Delta T$ of $T_c$ is
taken to be given by $\Delta T \propto n_T^{-4/3}$ [see  Eq. (11)], but we
also suppose that $\Delta T$ is proportional to the inverse of the
carrier concentration in order to take into account that phase slips
are not so easily formed if the carrier concentration rises.  Hence we
suppose that the reduction in $T_c$,  $\Delta T_2$, for the highest
resistance superconducting channel in the 0.3 $\mu$m films, with
$n_T=17$, is related to the corresponding quantity, $\Delta T_1$ for
50 $\mu$m films, with $n_T=32$, by
\begin{equation}
\Delta T_2 \approx 2.3 (c_Ba_T/c_{B2}a_T)\Delta T_1,
\end{equation}
where $c_{B2}$ is the linear carrier concentration in the 0.3 $\mu$m films.

After other parameters are determined, $c_{B2}a_T$ is varied so that
\begin{equation}
T_B-\Delta T_2 = 295 K
\end{equation}
for $n_T=17$.  The change in $c_Ba_T$ also gives rise to a change in
$p=M_L/M_T$ because it changes the potential involved in calculating
the transverse mass.

\vskip 2em
\noindent
{\bf 13. Numerical Results}
\newline
For our numerical work we first choose a value for the parameter
$\Delta T_1$.  $T_B$ in our model as a function of $n_T$ reaches a
shallow maximum as a function of $n_T$, and then declines slightly as
$n_T$ decreases further. (A qualitative discussion of why this slight
decline occurs has been given for bosons with a quadratic dispersion in
Ref. \cite{Ea98b}).  However, the results shown in Fig. 1 indicate that once
$T_c$ reaches room temperature it remains above that temperature as
$n_T$ increases further, at least up to very high $n_T$.  Hence, when
other parameters are approximately determined, we may obtain a lower
limit for $\Delta T_1$ from this condition.  For our final values of
parameters we find that $\Delta T_1 > 7 K$, and for our calculations we
take a value
\begin{equation}
\Delta T_1 = 15 {\rm K}.
\end{equation}
That $\Delta T_1$ is probably not much larger than this may be argued
by comparison with conclusions of Zhao for carbon nanotubes\cite{Zh03,Zh06}.  
He gives an example (see pp 46-48 of Ref.\cite{Zh06}) where
phase slips give rise to a resistance equal to about 0.4 $\times$ (the
normal-state resistance at a temperature of the order of 0.8 $\times$
the transition temperature without phase slips) for a number of
(spin-degenerate) channels equal to  54, 
corresponding to $N_T^2=108$ for our type of system,  where we suppose
there is no spin degeneracy.  Hence, if we assume that Eq. (11) is true,
for our case we might suppose that $\Delta T_1/T_B$ for $32^2=1024$
nanofilaments is slightly larger than $[(1024/108)^{-2/3}=0.223]\times 0.2
\approx$ 0.045, implying $\Delta T_1$, given by $\Delta T_1=
0.045\times (295K+\Delta T_1)$ 
is slightly larger than 14 K.  Of
course this is only an order-of-magnitude estimate because Eq. (11) and the
analysis of Zhao are based on a theory appropriate for materials on the
BCS side of the BCS-BEC transition, whereas we are assuming we are on
the opposite side.

Having determined $n_T$ for the highest-resistance superconducting
channels in section 3, and chosen the value of $\Delta T_1$ given by
Eq. (53), we next make use of the requirement that $T_c\approx 295 K$ for
the highest-resistance superconducting channels for 50 $\mu$m films and
results of sections 4 to 11 to  determine another eight parameters of
our theory approximately. These are $a_T$, $c_Ba_T$ for 50 $\mu$m films,
$a_T/\rho$, $E_b$, $M_L$, $p=M_L/M_T$, $\hbar \omega$ and $\mu_B$.
This gives nine approximate equations for eight parameters, and so one
might think that the parameters are overdetermined.  However, because
many of the relations between parameters are very approximate, we do
not have to worry about satisfying them exactly, and it turns out that
we can satisfy them to within our estimated accuracy for the
equations.

Our procedure is as follows.  We regard Eq. (16) of section 4 as exact, and
then minimise a weighted sum of moduli of differences between sides
of dimensionless quantities based on other equations, with weightings
based on our estimates of accuracy of our equations.  In order not to
cause computing time to be too long, we approximate the sums required
to produce $T_B$ given in section 2 by grouping terms in the sums over
$k$ in section 2 as equal to their mean values between two limits, the
differences between these limits increasing as $k$ increases.  For our
detailed calculations, we regard the numerical equation giving $T_c=295$
K for the highest-resistance superconducting channel for 50 $\mu$m
films as accurate to less than 1\%, but supplemented by errors of up to
about 1\% because of our approximate methods of evaluating the sums
over $k$ in equations of section 2.  We weight the differences in
moduli of calculated and fitted values of quantities $x$ divided by the
fitted values, i.e.
\begin{equation}
|(x_c-x)/x|
\end{equation}
by weightings according to our guesses of the accuracy of the
equations.  Taking the weighting for $|[(T_{cc}-295K)/295K]|$, where $T_{cc}$ 
is the calculated $T_c$ for the highest-resistance superconducting
channel for 50 $\mu$m films, as 1, we choose the following much smaller
weightings for similar quantities associated with $a_T$ given by
Eq. (21) (0.06), $j_c$ given by Eqs. (41) and (42) (0.015), $p=M_L/M_T$
given by the methods described in section 6 (0.03), $\hbar \omega$
based on Eq. (35) (0.1), $E_{pb}$ based on Eq. (38)
(0.03), $M_L$ based on Eq. (36) (0.2), and $\mu_B$ based on methods
described in section 11 (0.1).  There is some guessing as to the
appropriate weightings.  The weighting for the transverse bipolaron mass is
taken as fairly low because its calculation depends both on details of
Grigorov's variational method, and assumptions about how to convert the
masses deduced from his theory to masses for pairs.  The very small
weighting for $j_c$  is due to the uncertainty in the experimental
values of critical currents, the larger uncertainty arising from
conversion to a critical current density, the uncertainty to what
extent 295 K represents a "low $T$" for calculation of critical
currents, and the uncertainty about how much other parameters change
when large currents are passed.  The weighting for $E_{pb}$ is small
because for typical values of parameters the ratio of the upper and
lower limits is quite large.  For computational purposes, for
calculations of $p$ we approximated results for logarithms of
expressions as polynomials in relevant parameters, and for $\mu_B$ we
approximated by a different simple expression in the relevant
parameters, and used the polynomial forms  or other simple expressions
in our minimisation programs.  Since $a_T/\rho > 2$, but probably not
much greater, we only did calculations for $c_Ba_T$ (=0.259$a_T/\rho)$
between 0.5 and 0.7.  For the eight-parameter minimisation, we made use
of a program AMOEBA from a book\cite{Pr86}, based on a method
suggested by Nelder and Mead\cite{Ne65}.  For our computing we used
dimensionless lengths in units of nm, dimensionless masses in units of
$m_e$, and dimensionless energies in units of
$(\hbar^2/2M_L)(\pi/a_T)^2$.

From the best fit to the weighted sum of moduli as given by Eq. (54),
we find that
\begin{equation}
T_c=295.03\; {\rm K},\;\; a_T/\rho= 2.061, 
\end{equation}
implying with Eq. (16) that $c_Ba_T=0.534$,
and fitted and calculated values $x$ and $x_c$ of other quantities,  
and of $[(x_c-x)/x]$ found
for (i) $a_T$ from Eq. (21), (ii) $p$ from section 6, (iii) the plasmon
energy from Eq. (35), (iv) $M_L$ from Eq. (36), $E_{pb}$
from Eq. (38), (v) $j_c$ from Eqs. (41) and (42), (vi) $\mu_{B}$ from 
section 11 are:
\begin{eqnarray}
a_T=1.282 {\rm nm}, \;\;& (a_{Tc}-a_T)/a_T =0.033\nonumber\\
p=0.315,\;\;& (p_c-p)/p =0.002\nonumber\\
\hbar \omega = 1.31{\rm eV}\;\;& (\hbar \omega_c-\hbar \omega)/\hbar \omega =
 -1.3\times 10^{-6} \nonumber\\
M_L=2.376 m_e,\;\;& (M_{Lc}-M_L)/M_L=0.029\nonumber\\
E_{pb}=0.415 {\rm eV},\;\;&(E_{pbc}-E_{pb})/E_{pb}=7\times 10^{-5}\nonumber\\
j_c= 3.43\times 10^8 {\rm A cm}^{-2},\;\;& (j_{cc}-j_c)/j_c=-0.510
\nonumber\\
\mu_{B}= 0.00127 {\rm eV},\;\;& (\mu_{Bc}-\mu_B)/\mu_B=1.1\times 10^{-5}.
\end{eqnarray} 
The only quantity for which the calculated values and the values
obtained by our fitting process is significantly different is $j_c$.
Some reasons why we expected that the $j_c$ estimate would be
inaccurate are stated earlier in the section. 
For the values of parameters
found, the upper limit for $E_{pb}$ in section 9 is 0.79 eV and the
lower limit is 0.045 eV.  It turns out that our lower limit is smaller
than the alternative lower limit of 0.08 eV given in section 9, but if
we had used this alternative in our computation we do not expect that
there would be much change in parameter values.

Note that the small differences in most cases between calculated and
fitted parameters shown above do not imply that the parameters are
known to anything like that accuracy, as  errors cannot be expected to
be less than our estimated errors in the calculated relations used in
our fitting process.

After the other parameters are obtained, we treat $c_{B2}a_T$ for the
channels through 0.3 $\mu$m films as adjustable, where $c_{B2}$ is the
linear concentration of bosons in these channels.  The calculation of
transverse mass in this calculation has to be modified for the
different carrier concentration.   We find that $c_{B2}a_T=0.636$
(compared with $c_Ba_T=0.534$ for 50 $\mu$m films), and the the ratio
$p_2$ of longitudiinal to transverse mass in the channels through the
thinner films is given by $p_2=0.$258, smaller than the  value
$p=0.315$ for the thicker films.  Note that, if other parameters were
to remain fixed, then $T_B$ would increase with boson concentration.
However, $p=M_L/M_T$ decreases with increase of $c_B$ because of the
increase in depth of the potential of Eq. (25) with increasing $c_B$.
Thus there is a subtle competition between two effects in the parameter
range in which we are interested.

For the values of parameters found, $T_c=T_B-15$ K $\times
(n_T/32)^{-4/3}$  reaches a maximum  of about 303 K for 50 $\mu$m films
for $n_T\approx 140$, and $T_c=T_B-15$
K$\times(n_T/32)^{-4/3}(c_Ba_T/c_{B2}a_T)$ for 0.3 $\mu$m films reaches a
maximum of about 310 K for $n_T\approx 90$, assuming channels are
single domains with all spins aligned.  For relatively large channels
it is likely that the channels will split up into domains with opposite
directions of spins, but we do not know enough about the transverse
exchange constants or magnetic anisotropy to be able to predict the
maximum diameter of single-domain channels.  In order to explain higher
$T_c$'s than the two values noted above, observed when relatively high
currents are applied\cite{En89}, we can invoke decreases in transverse
masses due to reduced lattice constants produced by current-current
interactions between nanofilaments as discussed in some detail in
Ref. \cite{Ea05}.

\vskip 2em
\noindent
{\bf 14. Discussion}
\newline
In this section we discuss many of the assumptions that come into our model,
and to what extent they can be justified.   We also discuss some other 
matters.

\vskip 2em
\noindent
{\em Assumption of an array of nanofilaments}
\newline
There is at present no experimental evidence that the conducting
channels are composed of an array of nanofilaments, and so for this
assumption we have to rely on the plausibility of the theory of
Refs. \cite{Gr90a,Gr90b,Gr91b} for strings of an unusual type of polaron
occurring in elastomers with rotatable dipolar groups.   In this
connection we remark that there are also theories for strings of more
conventional polarons (Ref. \cite{Ku04} and references therein, 
also Ref. \cite{Al00}), although to form these one has to include interactions
other than the usual Fr\"ohlich interactions with polar phonons and
Coulomb repulsions, and there are limits to the lengths of the
strings\cite{Al00}.

\vskip 2em
\noindent
{\em Assumption of a square lattice}
\newline
We cannot rule out a triangular lattice.  Our use of a square lattice
is based on the simplicity of calculations, and the probability that
results for a triangular lattice  will not be  much
different, at least for weak or intermediate coupling of electrons and
bosons mediating the attraction.  For strong coupling a triangular
lattice is likely to be favoured because of the smaller transverse
masses of bipolarons on such a lattice when coupling becomes strong
\cite{Ha07b}.

\vskip 2em
\noindent
{\em Assumption that our system is on the BEC side of the BEC-BCS transition}
\newline
We estimate the pair diameter $d_p$ in the longitudinal direction from
the pair binding energy and reduced mass of the particles making up the
pair, and compare this with the separation between pairs $s\approx
(c_Ba_T)^{-1}a_T$.   We use the approximation sign rather than the
equality for $s$ because there will be a slight deviation from this
value towards the (lower) average separation $(c_Ba_T)^{-1/3}a_T$ which
would occur if the system were three-dimensional with no cut off in wave
vector in the transverse direction.  We find
$d_p=4(\hbar^2/2M_LE_{pb})^{1/2}=0.79$ nm, $s\approx 2.40$ nm, and so
the diameter of the pairs is considerably smaller than their mean
separation.  This shows that we are well on the BEC side of the transition.

\vskip 2em
\noindent
{\em Assumption of good contacts}
\newline
If the contacts were not good, then $n_T$ for the highest-resistance
superconducting channels would have to be larger for both thicknesses
of films, and then it would be more difficult to justify a $\Delta T_1$
as large as we need to  permit $T_c$ to increase monotonically with
$n_T$.  Also, from Fig. 1, the ratio of resistances of the
lowest-resistance mediumly conducting channels imply, taking into account the
different channel lengths, that the ratios of cross sections for the
lowest-resistance normal channels for the two film thicknesses needs to
be about four, the same as inferred for the highest-resistance superconducting
channels assuming good contacts.  Hence, if the contacts are not good,
they would have to have about the same departure from goodness for both
thicknesses of films, which would be another assumption.  For
discussion of normal channel resistances, although with smaller values
of $n_T$ than given by the present model, see Ref. \cite{Ea05}.

\vskip 2em
\noindent
{\em Assumption of a Bogoliubov-type of dispersion for bosons at $T_c$}
\newline
A possible justification for this assumption is discussed in the
Introduction.  We tried to fit the data assuming a simple quadratic
dispersion, but were unable to do this with plausible values of
parameters.  However, we could fit the data with a fixed value of $\mu_B$
[in units of $E_u=(\hbar^2/2M_L)(\pi/a_T)^2]$ over three orders of
magnitude smaller than that found by our data fitting.  For
$\mu_B/E_u=10^{-5}$ [cf. $\mu_B/E_u=0.0132$ for the results given in
section 13], $a_T$ is reduced to 1.037 nm, $p$ is increased to 0.423,
$M_L=2.444 m_e$, and $a_T/\rho=2.000$.  For $\mu_B/E_u$ significantly
smaller, $T_B$ for $n_T=32$ is too small for plausible values of $a_T$
and boson concentrations, e.g. for $\mu_B=0$, $L$=50 $\mu$m, $a_T=1$ nm,
$p=0.5$, $ca_T=0.54$, $M_L=2.55 m_e$, $T_B$ for $n_T$ =32 is only 25K,
compared with nearly 500K for large $n_T$.   The only possibility to
fit data with $\mu_B=0$ would be to assume poor contacts, so that the
lowest $n_T$ to give $T_c=295$K is considerably larger.  We tried
fitting assuming that $n_T$ for the highest-resistance superconducting
channel for the 50 $\mu$m films was $n_T=100$ (setting $\Delta T=0$ in
this case because of the large $n_T$), but did not get nearly so good a
fit as for our Bogoliubov-type dispersion and $n_T=32$.  For $n_T=100$,
we found that the calculated value of $a_T$ was 51\% higher than the
fitted value, the calculated value of $p$ was 72\% lower than the
fitted value, and the weighted sum of moduli of differences between
calculated and fitted values was about seven times higher than for our
previously reported results for $n_T=32$ and $\mu_B$ as a fitting
parameter. The problem with trying to fit with low values of $n_T$ for
$\mu_B=0$ is that, for quadratic dispersion, the rise of $T_B$ with $n_T$
is rather slow, and so, if we use Grigorov's expressions for linear
carrier concentrations $c_B$, it is not possible to get $T_B$ for small
$n_T$ up to room temperature unless $a_T$ is small, and then it becomes
impossible to satify Eq. (21), since we have the restriction
that $a_T/\rho>2$.  Another problem is the strong length dependence of
$T_B$ for quadratic dispersion (proportional to $1/L$ for small $n_T$),
and so one would need to have a considerable change in carrier
concentration in the 0.3 $\mu$m films to find a room-temperature $T_B$
for the highest-resistance superconducting channel with parameters
determined by fitting $T_B$ for the highest-resistance superconducting
channel for 50 $\mu$m films.  Since computation time for our program
goes up as the square of $n_T$, it is not practicable to see whether
things improve for pure quadratic dispersion if $n_T$ is taken as much
larger than 100, without making further approximations for $T_B$.  For
the rather drastic approximation made by Eq. (21) of Ref. \cite{Ea98b},
but increased in magnitude by 15\%, since, for
the cases we studied there gave errors of the order of 15\%
in $T_B$, we found that the best fit was obtained when $n_T$ for the
highest-resistance superconducting channel for the 50 $\mu$m film was about
240, with values of other parameters as follows:
\newline
$p=0.596$, $a_T/\rho=2.003$, $a_T=0.942$ nm, $M_L/m_e=2.444$,
$\hbar\omega=1.96$ eV, $E_b=0.621$ eV, $j_c=1.08\times 10^{9}$ A
cm$^{-2}$.  However, neither the goodness of the fit nor values of all
parameters vary monotonically with $n_T$, which may indicate that we do
not always reach the absolute minimum of our weighted sums with our
program.  For these values of parameters, $n_Ta_T$ for the highest
resistance superconducting channel for 50 $\mu$m films is
approximately equal to 230 nm.  Since there is a spread of resistance
for the low-resistance channels for these films shown in Fig. 1 of
about 100, this would imply that the widest channels shown in Fig. 1
would be about 2.3 $\mu$m in diameter, slightly larger than the largest
possible value mentioned in Ref. \cite{De90}.

We may look at results for pure quadratic dispersion in more detail
sometime, if we can find a better approximation to $T_B$ which does
not require too much computing time to investigate.  Although we think
that our arguments about fluctuations giving parts of the samples at
any one time with higher $T_B$'s than average, and hence a
Bogoliubov-type of dispersion for these parts is plausible, it is less clear
whether it is appropriate to consider some average type of Bogoliubov
dispersion over the whole sample with a lower effective value of $\mu_B$,
as we have done for most of this paper.

\vskip 2em
\noindent
{\em Remarks on Grigorov's theory for ferromagnetic nanofilaments}
\newline
It might be objected that a paper by Lieb and Mattis\cite{Li62} could
rule out ferromagnetism in quasi 1-D filaments.  However, on studying
this paper, it appeared to the present author that Lieb and Mattis's
result depended on the system being really one dimensional, not just
quasi one dimensional in the sense that all the carriers are in the
lowest state with respect to transverse motion,  and thus we think
their result is not relevant to any real system.  Grigorov\cite{Gr91b}
also notes that the result of Lieb and Mattis is based on a model with
short-range forces.  Further, Shelykh et al.\cite{Sh03} also find
ferromagnetism in  a quasi one-dimensional electron gas, and Meyer and
Matveev\cite{Me09} mention, in connection with quantum wires, that
"deviation from true one dimensionality may, in principle, give rise to
a spin-polarised ground state of the interacting electron system".

There are more general objections to any form of long range order in
quasi one-dimensional systems, but these will not be relevant for
arrays of nanofilaments.

\vskip 2em
\noindent
{\em Stability of electron strings}
\newline
Grigorov\cite{Gr91b} finds that the ferromagnetic electron string is
only stable if $R_0$, the radius from an excess electron up to which
all dipoles point away from the electron, satisfies $R_0>0.6$ nm.
For our parameters and equations in section 4 we find $R_0\approx 0.61$
nm.  Putting this value into Grigorov's Eq. (9),  with the second
term reduced by a factor of 0.8 since we are assuming
$k=\epsilon_h/\epsilon_s=0.8$ instead of unity as in his calculations,
and using $\epsilon_h=2.2$, we find $E_{min}=(0.96-0.61)=0.35$ eV.
However, in our model there is a reduction in energy due to polaron and
bipolaron binding energies caused by interaction with high-energy
bosons.   From our fitting we find that the bipolaron binding energy
$E_{pb}\approx 0.42$ eV, with $E_{pb}/\hbar \omega$=0.32. From
Ref. \cite{Ea07}, section 4, for $a=4$, corresponding to $\hbar
\omega/t=16$, this value of $E_{pb}/\hbar \omega$ probably corresponds
to $V^2\approx 0.29$.  Then for weak coupling and $a=4$, from equation
(46) of Ref. \cite{Ea07}, the single-polaron binding energy $E_{sb}=0.28
\hbar \omega  = 0.37$ eV.  Hence we get a total energy lowering per
electron of $E_{sb}+0.5E_{pb}=0.58$ eV, quite enough to stabilise the
system.

Note that the value of $R_0=0.61$ nm noted above that we find is
considerably smaller than the estimate that $R_0$ is greater than about
2 nm mentioned in Ref. \cite{Gr91b}, and so, if our parameters are
approximately correct, then we think Grigorov must have underestimated
the shear modulus $G$ in Eq. (12).  Other quantities in this equation are
probably known far better than $G$.

\vskip 2em
\noindent
{\em Alternative assumption for $\hbar \omega$}
\newline
There are high-frequency phonons in polypropylene with energies
$\hbar \omega = 0.37-0.39$ eV\cite{McD61}.  We have also looked at our
model on the assumption that phonons of energy 0.38 eV mediate the
attraction.  The best fit when minimising the weighted sum of the same
quantities as before except for the calculated and fitted frequency of
bosons, gives $a_T/\rho=2.055$, and the following values of other
parameters and departures from their calculated values:
\begin{eqnarray}
a_T= 1.322 {\rm nm},\;\;& (a_{Tc}-a_T)/a_T =-8\times 10^{-7}\nonumber\\
p=0.234,\;\;& (p_c-p)/p =9\times 10^{-7}\nonumber\\
M_L=2.527 m_e\;\;& (M_{Lc}-M_L)/M_L=-7\times 10^{-4}\nonumber\\
E_{pb}=0.143 {\rm eV},&\;\;(E_{pbc}-E_{pb})/E_{pb}=-0.063\nonumber\\
j_c=2.84\times 10^{8} {\rm A cm}^{-2}\;\;& (j_{cc}-j_c)/j_c=-0.595
\nonumber\\
\mu_{B}= 0.00263{\rm eV},\;\;& (\mu_{Bc}-\mu_B)/\mu_B=1.1\times 10^{-7}.
\end{eqnarray} 

Differences of calculated and fitted parameters are on
average not much different from values obtained assuming plasmons
mediate the interaction. 
For these parameters we find $d_p=1.29$
nm and $s\approx 2.48$ nm, and so we are still on the BEC side of the
transition, although not quite so far on this side as for our earlier
set of parameters.

\vskip 2em
\noindent
{\em Alternative theory of Grigorov for superconductivity}
\newline
Grigorov\cite{Gr98b,Gr99} has a completely different type of theory
for superconducting channels involving pairs of oppositely moving
charged density waves along each superpolaron string.  In his theory he
appears to think that the superconductivity will occur  even for
isolated superpolaron strings.  If that were the case we would have to
look for a different explanation for the difference between normal and
superconducting channels from that used here,  and for the difference
in resistances of the highest-resistance superconducting channels for the
two film thicknesses.

\vskip 2em
\noindent
{\em Remarks on use of one-dimensional Fermi wave vector}
\newline
Our model does not involve a truly one-dimensional system.  However,
for our fitted values of parameters, the transverse wave vector for a
given longitudinal wave number $k_L$ reaches its maximum possible value
$\pi/a_T$ for $k_L$ up to a value equal to a large fraction of the
calculated 1-D Fermi wave vector for a ferromagnetic system of
$k_{F,1-D}=\pi c_F= 2\pi c_B$, where $c_F$ and $c_B$ are the linear
concentrations of fermions and bosons.  Bearing this in mind, the
longitudinal Fermi wave vector $k_{FL}$ can be estimated as follows.  
First we approximate the square projection of the Brillouin zone on the
$xy$-plane by a circle of radius $k_c$ with the same area, i.e. of
radius given by $\pi k_c^2=(2\pi/a_T)^2$, or
\begin{equation}
k_c=2\pi^{1/2}/a_T.
\end{equation}
Then the maximum value of $k_m$ of $k$ at which the transverse wave vector
equals $\pi/a_T$ is given by
\begin{equation}
k_m=(k_{FL}^2- pk_c^2)^{1/2}.
\end{equation} 

If the longitudinal wave vector is $k_{FL}$, then the volume of the
Brillouin zone is the same as that for an area $\pi k_c^2=4(\pi/a_T)^2$
and length $2[(1/4)k_{FL} +(3/4)k_m]$.  The factor of 3/4 for the
weighting of $k_m$ occurs because of the quadratic dependence of the
relevant cross-sectional area on $(k_{FL}-k)$ up to $k=k_m$.  Since the
volume of the Brillouin zone $V_{BZ}=8\pi^3n_F =16\pi^3n_B$, where
$n_F$ and $n_B$ are the concentrations of fermions and bosons, we
deduce an equation for $k_{FL}$,
\begin{equation}
2[(1/4)k_{FL}+(3/4)(k_{FL}^2-4p\pi/a_T^2)^{1/2}]
(2\pi/a_T)^2=16\pi^3(c_Ba_T)/a_T^3.
\end{equation}

Putting in parameters from our model into these equations, we find that
$k_{F,1-D}=2\pi(c_Ba_T)/a_T=2.62$ nm$^{-1}$  and  $k_{FL}=2.95$ nm$^{-1}$,
i.e. the longitudinal Fermi wave vector is 13\% larger than the 1-D
value.  This is an error comparable with or smaller than most of our
estimated errors in calculated relations between various quantities in
our theory.  Note that, even if the longitudinal and transverse masses
were equal, the value of  the longitudinal Fermi wave vector would be
only 37\% larger than the 1-D value for the same values of other
parameters used, and so it is the cut off in transverse wave vectors
rather than the smaller transverse mass which is the main contributor
to the fact that the 1-D Fermi wave vector is a fair approximation.

\vskip 2em
\noindent
{\em Effect of replacing the probable tight-binding type of transverse
dispersion by one with a constant mass}
\newline
For a given mass at the bottom of a band, the single-particle bandwidth
in the $x$ or $y$ direction is larger by a factor of $\pi^2/4\approx
2.5$ for a band with a constant bare mass than for a tight-binding
band.  For a pair band made up from single-particle states with a
tight-binding dispersion, it is less obvious what happens to the
bandwidth.  If we form pair states from two single-particle states with
equal wave vectors (which may be a fair approximation for weakly
bound pairs), but do not consider pair states with transverse wave
vector greater than $\pi/a_T$ because of instabilities (cf.
\cite{Po05a,Po05b}), then the width in the $x$ and $y$ directions due
to a tight-binding model will only be decreased by a factor $(\pi^2/8)
\approx$ 1.23.  However, whatever happens to the pair bandwidth, if the
condensation temperature is such that $k_BT_B$ is large compared with
the transverse bandwidth, we expect the bandwidth is what is most
important, whereas for the opposite inequality the states near the
bottom of the transverse band are what matters.  For our parameters the
transverse bandwidth in the $x$ and $y$ directions is
$p(\hbar^2/2M_L)(\pi/a_T)^2$ = 0.0293 eV, and in the [110] direction
twice this.  Thus we expect that with a tight-binding model the
transverse mass at the bottom of the band to fit the data would be
somewhat smaller that determined by our data fitting, but probably not
much smaller.

\vskip 2em
\noindent
{\em Value of $\mu_B$ at $T_c$ compared with $\mu_{B0}$ at $T=0$.}
\newline
The value of the sound velocity $v_s$ at $T=0$ in the Bogoliubov theory
is 
\begin{equation}
v_s=(\mu_{B0}/M_B)^{1/2}, 
\end{equation}
where $\mu_{B0}$ is the bosonic chemical
potential at $T=0$ and $M_B$ is the boson mass (see e.g. Ref. \cite{Pi04}).
For the sound velocity along the channel direction we use the
longitudinal boson mass $M_L$, and we also note that, 
in the BEC r\'egime\cite{He06}
\begin{equation}
v_s=(0.165\pm 0.085)v_F 
\end{equation}
(see section 10), where the Fermi velocity  in the
longitudinal direction is given by
\begin{equation}
v_{FL}=\hbar k_{FL}/m_{FL}\approx 2\hbar k_{FL}/M_L.
\end{equation}
We deduce  from Eqs. (61) to (63), with $M_B=M_{L}$  and $v_F=v_{FL}$ that
\begin{equation}
\mu_{B0} \approx 0.22(\hbar^2 k_{FL}^2/2M_L).
\end{equation}
Using the 1-D value for $k_{FL}$ and putting in values from section 13
for $M_L$, $c_Ba_T=0.259(a_T/\rho)$ and $a_T$, we deduce that
$\mu_{B0} \approx 0.024$ eV.   Hence, with our value for $\mu_B$ at
$T_c$ of 0.00127 eV, we find $\mu_B/\mu_{B0}\approx 0.05$, with quite
a large uncertainty because of the uncertainties in Eq. (62).
Interpreting the above ratio as the fraction of fluctuating condensate
at $T=T_c$, we deduce that this fraction is about 5\%.  We do not have a
theory for this fraction at present, and so cannot say whether its
value is what should be expected.

\vskip 2em
\noindent
{\em Coincidence of maximum $T_c$ at low currents  being only slightly
above room temperature} 
\newline
Within our theory as presented, the only way to increase the maximum $T_c$
would be to increase $\Delta T_1$ to considerably higher values than the value
of 15 K assumed, and adjusting other parameters accordingly. 
This would require ignoring our estimated order-of-magnitude
value of $\Delta T_1$ based on comparison with analysis of Zhao
for carbon nanotubes assumed to be in the BCS-r\'egime.

\vskip 2em
\noindent
{\em Shortage of experimental input}
\newline
Although we have managed to determine twelve parameters in our theory
approximately, most of the information needed to determine them has
come from theory rather than from experiment.  The only use of
experimental data has been (i) the use of the histograms in Fig. 1 to
determine the probable value of the lowest $n_T$ for two film
thicknesses for which the superconducting $T_c$ reaches room
temperature, (ii) the very approximate estimate of carrier
concentration in the channels used in section 5, and (iii) the even
more approximate estimate of the critical current density given in
section 10.   There is further information from magnetic properties
which could be used in principle.  However, we have had to abandon a
theory for the metamagnetic transition given in Ref. \cite{Ea98a} because
recent work\cite{Ea07}  on bipolaron dispersion at intermediate
coupling in one dimension does not support the theory of great
enhancements of electron-electron attractions at high drift 
velocities\cite{Ea94} on which the theory was based.  The large diamagnetism in
some samples has been interpreted in terms of induced currents in
superconducting channels forming closed loops\cite{Ea02}, but this
interpretation does not help to determine any of our parameters, except
to give an alternative estimate of channel diameters to those given in 
\cite{De90}.  In Ref.\cite{Ea02} we estimated that diameters for channels
which formed closed loops in one sample must be greater than 1.52 $\mu$m.

\vskip 2em
\noindent
{\em Some suggested experiments}
\newline
The greatest need in this field is for more experimental groups to try
to reproduce all experimental results previously reported mainly by one
group, at least as far as claims of room-temperature superconductivity
are concerned, and to perform various new experiments.  Nine
suggestions  for experiments were given in section 3 of Ref. \cite{Ea98a}.
Two of these were mainly to test a model for metamagnetism given there
which we no longer believe,  and another suggestion may not be
practicable, but most of the others seem to be worthwhile.  I mention
these here more briefly than in Ref. \cite{Ea98a}, and also make a couple of
new suggestions.  Those previously suggested which still seem to be
worthwhile are: 1. Use micro-Hall probes near the sample surface to
measure the time and spatial dependence of magnetic fields near the
surface.  One property which would be determined from such measurements
is the direction of the magnetic moment in the channels in metamagnetic
samples; 2. Perform magnetisation measurements for fields parallel to
sample surfaces.  One might expect a decrease of magnetisation with
time in such fields as opposed to an increase seen in fields
perpendicular to surfaces, attributed to gradual alignment of magnetism
in channels in the direction of the film thickness\cite{Gr96}; 3.
Attempt to study the channels by optical microscopy at oblique
incidence in order to have a component of electric field parallel to
the channel directions, and with various wavelengths to try to find the
size of any energy gap in the electron spectrum; 4. Perform electrical
measurements in magnetic fields higher than those required to produce a
metamagnetic transition to see whether the high conductivity remains in
this region; 5. Perform magnetic measurements below the glass
transition temperature to see if
time dependent effects observed above this temperature\cite{Gr96} in
magnetic fields no longer occur at low $T$; 6. Perform electrical
measurements with films under tension between surfaces to see whether
this enables one to find conducting channels parallel to film
surfaces.

Another suggestion which would help to test our model, which involves
resistance mainly due to phase slips in the smaller-resistance mediumly
conducting channels in Fig. 1, would be first to determine histograms
of resistance for microcontacts on a suitable film at room temperature,
and then to look at the temperature dependence of resistance at points
with resistance near the high-resistance end of the range for
superconducting channels and at the low-resistance end of the range of
the mediumly resistant points.  If our interpretation, which involves
some resistance due to phase slips, is correct, then the
lower-resistance mediumly conducting points would become highly
conducting at lower temperatures, while  points at the
higher-resistance end of the low-resistance points would change to
medium resistances at temperatures  somewhere between 295 K and
$(295K+\Delta T)$, where $\Delta T$ corresponds to the change of
transition temperature due to phase slips in the highest-resistance
superconducting channels for whatever film thickness is being studied.
Quantitative analysis of such experiments would enable $\Delta T$ and
$T_c$ to be determined as a function of $n_T$, and so would remove our
requirement to make guesses at $\Delta T$.  It would also be useful to
determine the relation between size of channels and their resistance
(in two-probe measurements) if the size of the channels can be
determined e.g. by electron microscopy.  Although the experiments
suggested in this paragraph would be quite time comsuming, we think
they would go a long way towards establishing superconductivity at room
temperature, and whether our model for it is realistic.  High
resolution electron microsopy could possibly detect the nm-size
subsystems of conducting channels used in our model.

\vskip 2em
\noindent
{\em New theory needed}
\newline
Two types of theory which would be useful to develop are: (i) An
extension of the Ambegoakar-Langer-McCumber-Halperin theory to help
determine amounts of phase slips for Bose-gas superconductors; (ii)
Calculations of transverse exchange energies and any magnetic
anisotropy in our model.  The first of these types of theory is likely
to be difficult.  To calculate the transverse exchange energy using
Grigorov's theory\cite{Gr91b} and our parameters should be
straightforward, but calculations of magnetic anisotropy and sizes of
domains do not appear to be simple.

\vskip 2em
\noindent
{\bf 15. Conclusions}
\newline
We have used a model of Bose condensation of bosons in arrays of
nanofilaments to interpret some results on room-temperature
superconductivity in narrow channels through films of oxidised atactic
polypropylene.  The model makes use of a theory of Grigorov and
coworkers involving nanofilaments composed of strings of an unusual
type of polaron produced by interaction with rotatable dipolar groups in
an elastomer, and then assuming that the polarons in the strings bind to
form bipolarons due to mediation of high-frequency bosons. These bosons
are assumed to be plasmons for most of our calculations, although a fair
fit to experimental and theoretical constraints may also be obtained
assuming the bosons are high-energy phonons with energy 0.38 eV.  There
are several differences from a model previously used\cite{Ea05},
especially (i) a different method of estimating the numbers of
nanofilaments in the highest-resistance superconducting channels, and
(ii) a smaller slope of the initial linear term in the dispersion of
the bosons than previously used.  This slope is based on a Bogoliubov
form of dispersion for bosons at $T_c$ with a slope determined by
requiring that the initial change from the condensation temperature for
an ideal Bose gas as the boson concentration is increased is about the
same as determined by accurate published calculations of the effects of
interactions on $T_B$ of a Bose gas.  There are twelve parameters in
our model.  For one parameter related to changes in $T_c$ due to phase
slips we can only determine a lower limit, and we choose a value
slightly higher than this limit.  The other parameters can then be
determined approximately by use of twelve relations between parameters
of varying degrees of estimated accuracy, four of these involving
experimental input and eight based on theoretical relations between
parameters.  Some suggestions for further experiments are made, and
also for theory which it would be worthwhile to develop.

\vskip 2em
\noindent
{\bf Acknowledgments}
\newline
I wish to thank L.N. Grigorov for  discussions in person and by e-mail
at various times up to early 2007, and A.N. Ionov for correspondence,
and for some comments on a draft of the Introduction.


\vskip 2em
\noindent
\begin{thebibliography}{99}

{\rm\bibitem{Gr88}
L.N. Grigorov  and
S.G. Smirnova, 
Deposited Article No. 2381,
All-Union Institute for Scientific and Technological Information, 
23 March 1988, {\bf V} p 88

\bibitem{En89}
N.S. Enikolopyan, L.N. Grigorov  and S.G. Smirnova,
{\em Pis'ma Zh. Eksp. Teor. Fiz.} {\bf 49}, 326 (1989) [{\em JETP Lett.}
{\bf 49}, 371 (1989)]  

\bibitem{Ea05}
D.M. Eagles, {\em Phil. Mag.} {\bf 85}, 1931 (2005)

\bibitem{Ar90}
V.M. Arkhangorodski\u{i}, A.N. Ionov, V.M. Tuchkevich 
and I.S. Shlimak, {\em Pis'ma Zh. Eksp. Teor. Fiz.} {\bf 51}, 56 (1990)
[{\em JETP Lett.} {\bf 51}, 67 (1990)]

\bibitem{De90}
O.V. Demicheva, D.N. Rogachev, S.G. Smirnova, E.I. Shklyarova,
M.Yu. Yablokov, V.M. Andreev  and L.N. Grigorov, {\em Pis'ma Zh. Eksp. Teor.
Fiz.} {\bf 51}, 228 (1990) [{\em JETP Lett.} {\bf 51}, 258 (1990)] 

\bibitem{Gr91a}
L.N. Grigorov, O.V. Demicheva  and S.G. Smirnova, {\em Sverkhprovodimost'
(KIAE)} {\bf 4}, 399 (1991) [{\em Superconductivity, Phys. Chem. Tech.}
{\bf 4}, 345 (1991)]

\bibitem{Zh01} 
G.M. Zhao and Y.S. Wang, condmat/0111268 

\bibitem{Zh03}
G.M. Zhao, condmat/0307770 

\bibitem{Zh06}
G.M. Zhao, in {\em Trends in Nanotube Research}, ed. Delores A. Martin 
(Nova Science, New York, 2006), pp 39-75 (2006)

\bibitem{Zh08} 
G.M. Zhao and. P. Beeli, {\em Phys. Rev. B} {\bf 77}, 245433 (2008)

\bibitem{Ts99}
V.I. Tsebro, O.E. Omel'yanovski\u{i} and A.P. Moravski\u{i},
{\em Pis'ma Zh. Eksp. Teor. Fiz.} {\bf 70}, 457 (1999) 
[{\em JETP Lett.} {\bf 70}, 462 (1999)]

\bibitem{Ko00}
Y. Kopelevich, P. Esquinazi, J.H.S. Torre and S. Moehlecke, {\em J. Low
Temp. Phys.} {\bf 119}, 691 (2000)

\bibitem{Ko07}
Y. Kopelevich and P. Esquinazi, {\em J. Low Temp. Phys.} {\bf 146}, 629
(2007)

\bibitem{Dj01}
D. Djurek, Z. Meduni\'{c}, A. Tonejc  and M. Paljevi\'{c},
{\em Physica C} {\bf 351}, 78 (2001)

\bibitem{Dj08}
D. Djurek, condmat/0811.4352 

\bibitem{Dj11}
D. Djurek, {\em J. Supercond. Nov. Mag.} {\bf 24}, 199 (2011)

\bibitem{Sm88}
S.G. Smirnova, O.V. Demicheva  and L.N. Grigorov,
{\em Pis'ma Zh. Eksp. Teor. Fiz.} {\bf 48}, 212 (1988) [{\em JETP Lett.}
{\bf 48}, 231 (1988)]

\bibitem{Gr93}
L.N. Grigorov, D.N. Rogachev  and A.V. Kraev, {\em Vysokomol.
Soedin.} B {\bf 35}, 1921 (1993) [{\em Polymer Science} {\bf 35}, 1625 (1993)].

\bibitem{Gr96} L.N. Grigorov, T.V. Dorofeeva, A.V. Kraev, D.N, Rogachev,
O.V. Demicheva and E.I. Shklyarova, {\em Vysokomol. Soedin.} A {\bf 38},
2011 (1996) [Polymer Science A {\bf 38}, 1328 (1996)]

\bibitem{Ro00}
D.N. Rogachev  and L.N. Grigorov, {\em J. Supercond.} {\bf 13}, 947 (2000)

\bibitem{Ea02}
D.M. Eagles, {\em J. Supercond.} {\bf 15}, 243 (2002)

\bibitem{Io90}
A.N. Ionov  and V.M. Tuchkevich, {\em Pis'ma Zh. Tekh. Fiz.} {\bf 16} (15-16),
90 (1990) [{\em Sov. Tech. Phys. Lett.} {\bf 16}, 638, (1990)]

\bibitem{Io92}
A.N. Ionov, A.N. Lachinov, M.M. Rivkin and V.M. Tuchkevich, {\em Solid State
Commun.} {\bf 82}, 609 (1992)

\bibitem{Io00} 
A.N. Ionov and V.A. Zakrevski\u{i}, {\em Pis'ma Zh. Tekn.
Fiz.} {\bf 26} (20), 34 (2000) [{\em Tech. Phys. Lett.} {\bf 26}, 910 (2000)]

\bibitem{Io08} 
A.N. Ionov,  R. Rentzsch and M.N. Nikoleeva, {\em Phys. Stat. Sol.} (c) {\bf
5}, 730 (2008)

\bibitem{Re08}
R. Rentzsch and A.N. Ionov, {\em Phys. Stat. Sol.} (c) {\bf 5}, 735 (2008)

\bibitem{Sh98}
I. Shlimak and V. Martchenkov, {\em Solid State Commun.} {\bf 107},
443 (1998).

\bibitem{Ba95}
M. Baran, V.A. Beloshenko, V.P. D'yakonov, \'E.E. Zukov, A. Nabialek 
and R. Shimchak, {\em Fiz. Tverd. Tela} {\bf 37}, 3438 (1995) 
[{\em Sov. Phys. Solid State} {\bf 37}, 1889 (1995)]


\bibitem{Gra98}
P.M. Grant, {\em Physics Today}, May 1998, p17

\bibitem{Gr98}
L.N. Grigorov and K.P. Shambrook, US Patent No. 5777292 (1998)

\bibitem{Gr03} 
L.N. Grigorov and R.V. Talroze, US Patent No. 6563132 (2003)

\bibitem{Gr06} 
L.N. Grigorov and A. Krayev, US Patent No. 7014795 (2006)

\bibitem{Kr99}
A.V. Krayev, T.V. Dorofeeva, E.I. Shklyarova and L.N.
Grigorov, {\em 9th. CIMTEC - World Forum on New Materials, Florence, Italy,
14-19 June 1998}; Advances in Science and Engineering Technology, Vol.
23: Science and Engineering of HTC Superconductivity, edited by P. Vincennzini
(Faenza Techna., Srl., 1999), pp. 459-466.

\bibitem{Io02a} A.N. Ionov, A.N. Lachinov and R. Rentzsch, {\em Pis'ma Zh.
Tekh. Phys.} {\bf 28} (14), 69 (2002) [{\em Tech. Phys. Lett.} {\bf 26}, 608 
(2002)] 

\bibitem{Io02b} A.N. Ionov, V.A. Zakrevski\u{i}, V.M. Svetlichny  and
R. Rentzsch, {\em 10th. Int. Symp. "Nanostructures: Physics and Technology"},
St. Petersburg, Russia, June 17-21, 2002  [{\em SPIE Proceedings} 
{\bf 5023}, 475 (2003)]

\bibitem{Io99}
A.N. Ionov, V.A. Zakrevski\u{i} and I.M. Lazebnik, {\em Pis'ma Zh. Tekh. Phys.}
{\bf 25} (17), 36 (1999) [{\em Tech. Phys. Lett.} {\bf 25}, 691 (1999)] 

\bibitem{Io09a} A.N. Ionov and R. Rentzsch, {\em Annalen der Physik} 
{\bf 18}, 963 (2009)

\bibitem{La93} A.N. Lachinov, {\em Sensors and Actuators} A {\bf 39}, 1 (1993)

\bibitem{Sc95} O.A. Scaldin, O.A. Selezneva, Y.A. Lebedev,  A.N. Chuvrov,
{\em J. Appl.  Phys.} {\bf 77}, 3194 (1995)

\bibitem{La06} A.N. Lachinov, {\em Physics-Uspekhi} {\bf 49}, 1223 (2006)

\bibitem{Za98}
V.A. Zakrevski\u{i}, A.N. Ionov and A.N. Lachinov, {\em Pis'ma Zh. Tekh. Fiz.}
{\bf 24} (13), 89 (1998) [{\em Tech. Phys. Lett.} {\bf 24}, 539 (1998)]

\bibitem{Io10a} A.N. Ionov and R. Rentzsch, in {\em Proc. of SPIE 7521,
75210O (2010) - International Conference on Micro- and
Nano-Electronics 2009}.

\bibitem{Io10b} A.N. Ionov and R. Rentzsch,  to be published.

\bibitem{Io09b} A.N. Ionov, M.S. Dunaevskii, M.N. Nikolaeva and R. Rentzsch,
{\em Annalen der Physik} {\bf 18}, 959 (2009)

\bibitem{Kr93} A.V. Kraev, S.G. Smirnova and L.N. Grigorov, 
{\em Vysokomol. Soedin.} A {\bf 35}. 1308 (1993) [{\em Polymer Science} 
{\bf 35}, 1082 (1993)]

\bibitem{Gr90a}
L.N. Grigorov, {\em Makromol. Chem., Macromol. Symp.}
{\bf 37} 159 (1990) 

\bibitem{Gr90b}
L.N. Grigorov, V.M. Andreev and S.G. Smirnova, {\em Makromol.
Chem., Macromol. Symp.} {\bf 37} 177 (1990) 

\bibitem{Gr91b}
L.N. Grigorov, {\em Pis'ma Zh. Tekh. Fiz.} {\bf
17} (5), 45 (1991)  [{\em Sov. Tech. Phys. Lett.} {\bf 17} 368 (1991)]

\bibitem{Gr98b}
L.N. Grigorov, {\em Phil. Mag. B} {\bf 78}, 353 (1998).

\bibitem{Gr99}
L.N. Grigorov,  {\em 9th. CIMTEC - World Forum on
New Materials, Florence, Italy, 14-19 June 1998}; Advances in Science
and Engineering Technology, Vol. 23: Science and Engineering of HTC
Superconductivity, edited by P.Vincennzini (Faenza Techna. Srl., 1999), pp.
675-684

\bibitem{Ea94}
D.M. Eagles, {\em Physica C} {\bf 225}, 222 (1994) ; erratum 
{\em ibid.} {\bf 280}, 335 (1997) 

\bibitem{Ea98a}
D.M. Eagles, {\em J. Supercond.} {\bf 11}, 189 (1998)  

\bibitem{Ea07}
D.M. Eagles, R.M. Quick and B. Schauer, {\em Phys Rev.} B {\bf 75}, 054305
(2007)

\bibitem{Fe04}
M. Ferrier, A. de Martino, A. Kasumov, S. Gu\'{e}ron, M. Kociak, R. Egger
and H. Bouchiat,  {\em Solid State Comm.} {\bf 131}, 615 (2004)

\bibitem{Ad00}
S.K. Adhikari, M. Casas, A. Puente, A. Rigo, M. Fortes, M.A. Sol\'{i}s,
M. de Llano, A.A. Valladares and O. Rojo, {\em Phys. Rev. B} {\bf 62},
8671 (2000)

\bibitem{Ar01}
P. Arnold and G. Moore, {\em Phys. Rev. Lett.} {\bf 87}, 120401 (2001)

\bibitem{Ka01}
V.A. Kashurnikov, N.V. Prokof'ev and B.V. Svistunov, {\em Phys. Rev. Lett.}
{\bf 87}, 120402 (2001)

\bibitem{Pi08}
S. Pilati, S. Giorgini and N. Prokof'ev, {\em Phys. Rev. Lett.} {\bf 100},
140405 (2008)

\bibitem{Az68}
L.G. Aslamozov and A. Larkin, {\em Fiz. Tverd. Tela} {\bf 10}, 1104 (1968)
[{\em Sov. Phys. Solid State} {\bf 10}, 875 (1968)]

\bibitem{Al93} A.S. Alexandrov and N.F. Mott, {\em Phys. Rev. Lett.} {\bf 71},
1075 (1993)

\bibitem{Io05} A.N. Ionov, V.M. Svetlichinyi and R. Rentzsch, {\em Physica B} 
{\bf 359-361}, 506 (2005)

\bibitem{Lo80} J. Lowell and A.C. Rose-Innes, {\em Adv. Phys.} {\bf 29},
947 (1980)

\bibitem{Ea98b} D.M. Eagles, {\em Physica C} {\bf 301}, 165 (1998)

\bibitem{Al01} V.A.Alekseev, {\em Zh. \'{E}ksp. Teor. Fiz.} 
{\bf 119}, 700 (2001) [{\em JETP} {\bf 92}, 60 (2001)]

\bibitem{Al02} V.A. Alekseev,  {\em Zh. \'{E}ksp. Teor. Fiz.} 
{\bf 121}, 1273 (2002) [{\em JETP} {\bf 94}, 1091 (2002)]

\bibitem{Sch99} C, Sch\"{o}nenberger, A. Bachtold, C. Strunk, J.-P. Salvetat
and L. Forro, {\em Appl. Phys. A} {\bf 69}, 283 (1999) 

\bibitem{La67} J.S. Langer and V. Ambegoakar, {\em Phys. Rev.} {\bf 164}, 
498 (1967)

\bibitem{McC70} D.E. McCumber and B.I. Halperin, {\em Phys. Rev.} B 
{\bf 1}, 1054 (1970)

\bibitem{Po05a} A. Polkovnikov, E. Altman, E. Demler, B.I. Halperin and 
M.D. Lukin, {\em Phys. Rev.} A {\bf 71}, 063613 (2005)

\bibitem{Po05b} A. Polkovnikov, E. Altman, E. Demler, B.I. Halperin and 
M.D. Lukin, {\em J. Supercond.} {\bf 17}, 577 (2005)

\bibitem{Sh03} I.A. Shelykh, N.T. Bagraev, V.K. Ivanov and L.E. Klyachkin,
{\em J. Supercond.} {\bf 16}, 355 (2003)

\bibitem{Sm89} S.G. Smirnova, E.I. Shklyarova and L.N. Grigorov,
{\em Vysokomol. Soedin} B {\bf 31}, 667 (1989)

\bibitem{Al08} A. Alvermann, H, Fehske and S.A. Trugman, {\em Phys. Rev.} B
{\bf 78}, 165106 (2008)

\bibitem{Ha07} J.P. Hague, P.E. Kornilovitch, A.S. Alexandrov and 
J.H. Samson,{\em  Physica} C {\bf 460-462}, 1115 (2007)

\bibitem{Ha09} J.P. Hague and P.E. Kornilovitch, {\em Phys. Rev.} B {\bf 80}, 
054301 (2009)

\bibitem{Ho05} M. Hohenadler and W. von der Linden,  {\em Phys. Rev.} B 
{\bf 71}, 184309 (2005)

\bibitem{Pi05} P. Pieri, L. Pisani and G.C. Strinati, {\em Phys. Rev.} B 
{\bf 72}, 012506 (2005) 

\bibitem{Ch05} C. Chin, {\em Phys. Rev.} A {\bf 72}, 041601 (2005)

\bibitem{Je06} L.M. Jensen, H.M. Nilsen and G. Watanabe, {\em Phys. Rev.} A 
{\bf 74}, 043608 (2006)

\bibitem{Ea69} D.M. Eagles, {\em Phys. Rev.} {\bf 186}, 456 (1969)

\bibitem{De90b}
O.V. Demicheva, D.N. Rogachev, 
V.M. Andreev, E.I. Shklyarova, S.G. Smirnova and L.N. Grigorov, 
{\em Vysokomol. Soedin.} {\bf 32} (1), 4 (1990) 

\bibitem{He06} H. Heiselberg, {\em Phys. Rev.} A {\bf 73}, 013607 (2006)

\bibitem{Pe04} D.S. Petrov, C. Salomon and G.V. Shlyapnikov, {\em Phys. Rev.
Lett.} {\bf 93}, 090404 (2004)

\bibitem{Pi04} P. Pieri, L. Pisani and G.C. Strinati, {\em Phys. Rev.} B {\bf
70}, 094508 (2004)

\bibitem{Pr86} W.H. Press, B.P. Flannery, S.A. Teukolsky and W.T. Vetterling,
{\em Numerical Recipes}, Cambridge University Press (1986)

\bibitem{Ne65} J.A. Nelder and R. Mead, {\em Computer Journal} 
{\bf 7}, 308 (1965)

\bibitem{Ku04} F.V. Kusmartsev, {\em Contemp. Phys.} {\bf 45}, 237 (2004) 

\bibitem{Al00} A.S. Alexandrov and V.V. Kabanov, {\em Pis'ma Zh.Eksp.Teor.Fiz.} 
{\bf 72}, 825 (2000) [{\em JETP Lett.} {\bf 72}, 569 (2000)]

\bibitem{Ha07b} J.P. Hague, P.E. Kornilovitch, J.H. Samson and A.S. Alexandrov, 
{\em Phys. Rev. Lett.} {\bf 98}, 037002 (2007)

\bibitem{Li62} E. Lieb and D. Mattis, {\em Phys. Rev.} {\bf 125}, 164 (1962)

\bibitem{Me09} J.S. Meyer and K.A. Matveev, {\em J. Phys.: Cond. Mat.} 
{\bf 21}, 023203 (2009)

\bibitem{McD61} M.P. McDonald and I.M. Ward, {\em Polymer} {\bf 2}, 
241 (1961)
}
\end{thebibliography}
\end{document}